# Assessing the Feasibility, and Efficacy of Virtual Reality Navigational Training for Older Adults

**Tong Bill Xu** [1]

**Armin Mostafavi** [1]

**Walter R. Boot** [2]

**Sara Czaja** [2]

**Saleh Kalantari** [1*]

[1] Human Centered Design Department, Cornell University

[2] Division of Geriatrics and Palliative Medicine, Weill Cornell Medicine

[*] Corresponding Author: Saleh Kalantari <sk3268@cornell.edu>; 2427 Martha Van Rensselaer Hall, Ithaca, NY 14853

## Abstract

**Objective.** Evaluate the feasibility of Virtual Reality (VR) wayfinding training with aging adults, and examine the impact of the training on wayfinding performance.

**Design.** Design involved wayfinding tasks in a study with three groups: active VR training, passive video training, and no training, assigned randomly. The training featured 5 tasks in a digital version of a real building. Post-training assessments had 10 tasks in this building, half familiar from training and half new. The study was double-blinded, with each intervention lasting 10 minutes.

**Participants.** A convenience sample of 49 participants; inclusion criteria: age >58, unfamiliar with the building; exclusion criteria: mobility or vision impairments, history of motion sickness, or medical implants.

**Outcomes.** Time spent and Distance traveled on each wayfinding task with a fixed 10-min limit.

**Results.** Participants in VR group reported moderate usability (63.82, SD=14.55) with respect to the training intervention and high Self Location (3.71, SD=0.94). There were no differences in task performance among the groups in the similar tasks. In the new tasks, compared to the control condition, Time spent on tasks was marginally significantly reduced in the VR group; Distance traveled to finish tasks was also reduced in the VR group, and marginally significantly reduced in the Video training group. No differences were found between VR and Video conditions. No adverse effects were reported during or post intervention.

**Conclusions.** This study provides preliminary evidence that VR training can effectively improve wayfinding performance in older adults with no reported adverse effect.

***Keywords:*** Virtual Reality; Wayfinding; Spatial Navigation; Older Adults; Feasibility

## Translational Significance Statement:

This study addresses the challenge of declining wayfinding abilities in aging adults, a critical issue that can negatively affect quality of life. By evaluating the feasibility and effectiveness of Virtual Reality (VR) wayfinding training, we demonstrate preliminary evidence that VR can improve wayfinding performance in older adults without adverse effects. Implications for translation include the potential to improve individual autonomy and safety, reduce the risk of getting lost, and ultimately enhance the quality of life for aging populations by incorporating VR training into routine care or rehabilitation programs.

# 1. Introduction

The United States is expected to see a considerable shift in its age demographics by 2030, when over 70 million individuals, or roughly one-fifth of the total population, will be over the age of 65 (Vespa et al., 2020). Health concerns linked to older ages, such as Mild Cognitive Impairment (MCI), will continue to increase in prevalence due to this demographic shift. Current data suggests that the incidence of MCI could climb from its current level of about 22 per 1,000 individuals to well over 60 per 1,000 individuals (Gillis et al., 2019). Research has shown that older adults tend to under-perform in wayfinding and route-learning tasks and struggle with reading signage and acquiring configural knowledge when compared to their younger counterparts (Aubrey et al., 1994; Head & Isom, 2010), and that these wayfinding performance challenges are  exacerbated among those with MCI or Alzheimer's Disease and Related Dementias (ADRD) (Davis et al., 2017). The declines in wayfinding abilities among older adults can negative impact their overall quality of life (Taillade et al., 2013).

Digital games have been suggested as a promising avenue for mitigating some of these age-related deficits and helping older adults maintain their navigational skills. Prior studies have found that computer-based training or games may enhance cognitive functions associated with wayfinding (Green & Bavelier, 2003; Law et al., 2022), and possibly motor skills (Castel et al., 2005). In a more general sense, studies suggest that  game-like interventions can enhance cognitive functions or manage cognitive impairment in older adults (Kelly et al., 2021), and promote general learning (Green & Bavelier, 2012). However, findings regarding transfer of learning, the ability to apply skills learned in a gaming or training context to real-world scenarios (Barnett & Ceci, 2002), has been limited in studies of "brain training" interventions (Owen et al., 2010).

Virtual Reality (VR) can simulate complex real-world environments in a controlled setting. Research has shown that VR trainings improve cognitive functions (Anguera et al., 2013; Wais et al., 2021), physical functions (Molina et al., 2014) in older adults, as well as spatial learning (Meade et al., 2019; Parong et al., 2020). Further, research has shown that  3D models are more helpful than 2D floorplans for wayfinding (Verghote et al., 2019), and suggests that VR could have an advantage over desktop programs in terms of transferring of navigational skills to the real world (Hejtmanek et al., 2020).

In this study, we designed and developed a VR wayfinding training platform specifically for older adults, with realistic, task-based simulations (rather than artificial training exercises), to improve wayfinding skills. We aimed to evaluate the feasibility of the VR training, and its effects on wayfinding performance, wayfinding experience, and spatial learning outcomes.

# 2. Methods

## 2.1. Experiment Design

The study was a randomized design with the intervention training condition (VR, Video, Control) as the between study variable. The participants were assigned randomly into one of three intervention groups (VR, Video, or Control, 1:1:1) according to a randomization list generated in the R language.

The study was double-blinded, neither the experimenters or the participants were aware of the exact purpose of the study or its hypotheses. The experimenters were, however, aware of the difference in interventions in order for them to be able to conduct the study.

## 2.2. Apparatus

The virtual interior environment that the researchers created for navigational training was based on Martha Van Renslier Hall at Cornell University Ithaca campus. Most of the modeling and UV mapping for this environment was conducted using Autodesk 3ds Max. Texturing, lighting, and interactivity were added using the Unreal Engine 4.27. All of the front-end interaction and user interactivity leveraged the Blueprint platform and C++ scripting. The VR environment, as well as the videos watched by the Video and Control groups, were run on a Dell Alienware computer equipped with Nvidia RTX 2080 Ti to minimize the risk of latency. The VR environment was presented to participants using a cable-connected Meta Quest 2 head-mounted display, at a resolution of 1832×1920 pixels per eye (90 Hz, FOV 90°). Video for the control Participants in the Video and Control groups were seated in the same position while observing the relevant content on a 24-inch monitor (1920×1080 resolution, 60 Hz).

Participants experienced the VR environment from a seated position at a desk, and were able to look around freely to see various aspects of their virtual surroundings while using the controller to move (teleport) through the environment (Figure 1). They started tasks at the start location and were asked to find the end location by a prompt in VR, with guidance lines on the floor (Fig. 1D-F). Videos for Video group were rendered in Unreal with camera moving along the shortest path from start location to end location (Fig. 1A-C). Video used for Control group was a TED talk "Designing for virtual reality and the impact on education" by Alex Faaborg (TEDx Talks, 2015).

## 2.3. Participant Recruitment

A convenience sampling approach was used to recruit the older adult participants residing in Tompkins County, New York. This involved the distribution of fliers and posters in local retirement communities and sending e-mails to senior community mailing lists. To be eligible, individuals needed to be aged 58 or above and unfamiliar with the building used for wayfinding tasks in the experiment. Participants were excluded if they had substantial mobility issues, visual impairments, a history of epilepsy or motion sickness, medical implants, or scored below the recommended dementia threshold of 19 on the Montreal Cognitive Assessment (MoCA) (Milani et al., 2018) were excluded.

## 2.4. Outcomes and Measures

The details of all measurement tools were provided in Table 1. The baseline measurements included Sense of Direction, Cognitive Assessment, Computer Proficiency, and Mobile Device Proficiency. The primary outcome variable was Wayfinding Performance operationalized as Time Spent and Distance Traveled on each wayfinding task with a fixed 10-min limit. The researchers developed a Python-based application to track participant trajectories through the real-world building over time, which was used to measure the Distance Traveled **[citation to the software tool is removed for the purpose of blind review].**

Secondary outcomes were wayfinding experience, including Spatial Anxiety and Workload, and Spatial Learning, evaluated via two commonly used metrics: after completing each of the real-world wayfinding tasks, participants were asked to point in the straight-line direction where they believed the task's starting point was located (Pointing Error), and to estimate the straight-line distance to that origin point (Distance Error). We also included two additional intervention feasibility measures: Spatial Presence (Self Location and Possible Action), and User Experience. All primary and secondary outcomes were measured after completion of the training.

**2.5. Procedure**

The study took place in Ithaca, NY from January 2023 to June 2023. Prior to any research activities the study protocol was evaluated and approved by the Institutional Review Board (IRB) at Cornell University, and informed written consent was obtained from all participants. Experiment sessions were held for one participant at a time and took place in Martha Van Rensller Hall at Cornell University. Participants were asked to complete a demographic survey online before attending their experiment session; then upon their arrival they completed the MoCA test. A de-identified dataset and materials related to this study are available at https://osf.io/63t5z/. This repository has been established in line with ethical and legal guidelines to facilitate further research while ensuring participant confidentiality and privacy.

To ensure similarity between activities prior to wayfinding tasks across conditions, all participants completed a training session about teleportation interaction in VR without wayfinding element with headset on before the intervention. Next, the participants engaged in the VR training which involved viewing five real-world wayfinding tasks in the VR group, watching a wayfinding video covering same tasks in the Video Group, or watching an unrelated video in the Control Group,  (see Appendix 1 for details). The duration of the training session was 10 minutes across all conditions. Participants were not informed about the assessment tasks during training. After completing the training, participants were asked to complete the feasibility measures (see Spatial Presence, and User Experience in Table 1).

The participants then completed two sets of wayfinding activities, each containing five tasks. In each task, they started from one location and were asked to find another location in the building. The first five real-world tasks (*similar tasks*) had the same start and end locations of the wayfinding tasks included  in the VR/Video training sessions. The second five real-world tasks (*new tasks*) took place in a different part of the building that was not included in training.

Each set of wayfinding tasks formed a loop (Table S1, Figure S2), and each participant was assigned to a random starting point within the loop For example, one participant might start with Task 3 and from there complete, in order, Tasks 4, 5, 1, and 2. Each wayfinding task was designed to take approximately 3–6 minutes to complete; if a participant did not finish a task within 10 minutes then the researchers stopped the data-collection for that task and led the participant to the destination. This cutoff minimized the risk of informative right censoring (Shih, 2002; Templeton et al., 2020), and participant fatigue. A procedure flowchart was provided in figure S1.

### 2.6. Statistical Analysis

An a-priori power analysis was conducted to determine the required sample size with G*Power. Comparisons of primary and secondary outcome measures between three intervention conditions in the form of f-tests at 0.05 significance level required 159 samples (tasks) to reach 0.80 power, assuming a medium effect size (f=0.25). Taking the study design into consideration, with 5 tasks per cluster (participant) and assuming an intra-cluster correlation of 0.05, a sample size of 39 participants (13 participants per condition) was needed to achieve the required power. To account for a 10% attrition rate, we included 14 participants per condition, or 42 in total.

The data were analyzed at task level using the R statistical programming language (R Core Team, 2023). Data from tasks 1–5 (similar tasks) and from tasks 6–10 (new tasks) were analyzed separately. For the primary outcomes (Duration, Distance), we fitted mixed-effect Cox regression models (library "coxme") with fixed effects of intervention condition, adjusted for the fixed effects of task order, Sense of Direction (Hegarty et al., 2002) and the random effects (intercept) of participant and task, followed by $\chi^2$ tests and BIC-estimated BF10 (library "bayestestR").

For secondary outcomes, we fitted linear mixed models with same predictors, followed by f-tests and effect size estimation (library "effectsize"). We then estimated the Hazard Ratio (for primary outcome) and mean differences (secondary outcome) between intervention conditions and 95% CIs using the fitted models (library "emmeans"). We controlled for Sense of Direction due to its direct connection to wayfinding ability (Hegarty et al., 2002).

Additionally, we tested for the differences between groups in demographic/baseline and feasibility measures with f- or Fisher's exact tests. Test results were reported as significant if p-values were below 0.05, and as marginally significant if p-values were below 0.10. No subgroup analysis was prespecified/performed.

## 3. Results

### 3.1 Participants

We recruited a total 49 participants from January 2023 to June 2023. Five participants asked to leave before finishing all of the tasks: 2 participants in VR group, 1 in Video group, and 1 in Control group did not finish the last 5 tasks, 1 in Video did not finish the last 3 tasks. Two due to their schedules, three without giving specific reasons. In addition, we excluded 28 invalid trajectories due to incorrect start/end points, result in missing data in Distance measure (Table 3). No participant reported any adverse effects during training or the data collection session.

Demographic characteristics of randomized participants were similar across all groups (Table 2). The average age of the sample was 71.31 (SD = 7.68); 38 (78%) reported as Female, 11 (22%) reported as Male, and none reported as Other.

### 3.2 Primary Outcomes: Wayfinding Performance

Wayfinding Performance per group is reported in Table 3 and illustrated in Figure 2.

In the similar tasks, participants in the VR group spent an average of 262.67 seconds and traveled 174.69 meters, those in the video group spent 241.90 seconds and traveled 189.81 meters, while those in the control group spent 235.16 seconds and traveled 177.99 meters. There was no significant difference between groups in time needed to finish the tasks, $\chi^2(2) = 3.07$, $p = 0.215$, with anecdotal evidence $BF_{10} = 1.96$; or Distance traveled to finish the tasks, $\chi^2(2) = 3.78$, $p = 0.151$, with moderate evidence, $BF_{10} = 3.16$.

In the new tasks, participants in the VR group spent an average of 190.71 seconds and traveled 159.68 meters, those in the video group spent 203.38 seconds and traveled 166.92 meters, while those in the control group spent 234.93 seconds and traveled 206.05 meters.

We found a marginally significant intervention effect on time needed to finish the tasks, $\chi^2(2) = 5.26$, $p = 0.072$, and very strong evidence, $BF_{10} = 49.88$. Those in the VR group had a marginally significantly higher rate of finishing the tasks (HR = 1.71, 95% CI: [0.99, 2.96]), while the video group had no significant difference as compared to control group (HR = 1.45, 95% CI: [0.84, 2.52]). The difference between VR and video group was not significant, (HR = 1.18, 95% CI: [0.68, 2.05]).

The between group difference in Distance Traveled to finish the tasks was significant, $\chi^2(2) = 9.00$, $p=.011$, with extreme evidence, $BF_{10} = 439.95$. Those in the VR group had a significantly higher rate of finishing the tasks (HR = 2.03, 95% CI: [1.15,3.60]) than those in the control group, while those in the video group had a marginally significant difference (HR = 1.72, 95% CI: [0.96, 3.08]). The difference in Distance Traveled to finish the tasks between the VR and video group was not significant (HR = 1.18, 95% CI: [0.66, 2.11]).

### 3.3 Secondary Outcome Measures

There was no significant difference across intervention conditions in secondary outcomes, except for Workload in the similar tasks (Table 3, 4). Workload in the similar tasks was significantly different between groups, $F(2, 45) = 3.44$, $p=.041$, with a medium effect size, $\eta^2=0.13$, $\omega^2=0.09$. Those in the VR group reported significantly higher Workload as compared to those in the control condition, 0.71 (95% CI: [0.01, 1.40]). Workload did not differ between the video group and the control group (0.57, 95% CI: [0.13-1.28]), or the VR group (-0.14, 95% CI: [-0.85, 0.58]).

### 3.4 Feasibility Measures

For those in the VR training condition, the average usability ratings was 63.82 (SD=14.55) and the average rating of motion sickness was 13.86 (SD=16.71). Motion sickness and the Self Location sub-scale of Spatial Presence were significantly different between groups (Table 2). Self Location scores of the VR group were significantly higher than those of the Control group by 1.30 (95% CI: [0.23, 2.37); and marginally significantly higher than those of the Video group by 0.99 (95% CI: [-0.09, 2.06]). Participants in the Video group reported more motion sickness than those in the Control group by 20.10 (95% CI: [1.24, 38,96]) and marginally more motion sickness than those in the VR group by 17.23 (95% CI: [-1.35, 35.81]).

## 4. Discussion

In this study, we examined the effects of a VR wayfinding training intervention on subsequent real-world wayfinding performance, spatial learning, and wayfinding experience, in an older adult population. The study moved beyond the common lab settings and evaluated the intervention effects in a real and functional building.

In terms of feasibility, no participant reported any major adverse effects . Participants also reported a good user experience for the VR training environment, with moderate-good usability ratings (Bangor et al., 2009)and high Spatial Presence (Self Location subscale), which reflected the sensation of being within the virtual environment (Böcking et al., 2004). They also reported low levels of motion sickness, in fact lower than that of the video group.

Overall, we found that the VR training improved wayfinding performance only for the new tasks, those that were not included in training. This might indicate a greater amount of spatial learning in the VR group. The findings also suggested that the video training was effective, but only with marginal significance. We also found that the participants perceived that workload was higher in the the VR training as compared to workload ratings for those in the control condition for similar tasks. We did not find any significant differences between VR training and Video training in any of the secondary outcome measures, but note that the post-hoc comparisons/tests were exploratory and were not prespecified or considered in the a priori power analysis.

The effects of the VR training on wayfinding performance are similar to previous comparisons of VR and 2D wayfinding training, where VR was suggested to have "some advantage" over 2D training, but without significant difference between conditions (Hejtmanek et al., 2020). The delayed benefit of VR training was also reported in (Verghote et al., 2019), where the wayfinding performance difference between 2D floorplan and 3D model conditions was found in the second, but not the first half of the study. While our VR intervention was designed to simulate actual wayfinding tasks, rather than training isolated cognitive functions, the effects of the VR training as compared to the control condition on wayfinding performance had about a medium to large effect size (Azuero, 2016), close to that of cognitive training programs (Kelly et al., 2021).

The VR group's better performance in the second half of the tasks may suggest a higher level of spatial learning during the initial tasks. This was also suggested by the ratings of higher workload among those in the VR group. It is possible that the VR training intervention promoted more spatial learning (Meade et al., 2019; Parong et al., 2020), similar to what had been found for action video games (Green & Bavelier, 2012). Although the VR training intervention was not specifically created as a game and lacked some key elements recommended for fostering engagement and the appropriate level of cognitive load (Pasqualotto et al., 2023), it probably enhanced engagement through the sense of spatial presence (Kalantari et al., 2023) and induced cognitive load through tasks related to navigation and the use of teleportation controls.

Despite the difference in wayfinding performance, we did not find any differences between groups in the spatial learning measures. During the experiment, our participants remarked that while they could lead us to the task origin, they could not point at its direction, nor estimate their distance from it. Similarly, an early study found that student nurses who spent two years in a hospital were able to find various rooms/locations, but could not draw, or even understand the floorplan, and had poor pointing and distance estimation (Moeser, 1988). It is probable that the current methods used to measure spatial learning actually assess participants' capacity to form cognitive maps resembling floorplans (Kitchin, 1994), even though such maps are not always essential for navigating interior spaces.

### 4.1. Limitations and Future Studies

Our convenience sample was not representative of the overall older adult population; it skewed heavily toward female and white participants, with high levels of Computer Proficiency and Mobile Device Proficiency compared to previous studies (Boot et al., 2015; Roque & Boot, 2018). Future studies could benefit from applying a more systematic sampling method to improve participant diversity, to better assess the feasibility of the intervention. Additionally, while the self-reported Sense of Direction was similar to older adults recruited in other studies (Gandhi et al., 2021), future research should recruit more participants with lower Sense of Direction and consider subgroup analysis based on navigational abilities to examine the intervention effect on those who might benefit more from such interventions (Verghote et al., 2019).

Future studies should also explore different choice of design elements in the VR intervention. Some notable considerations in this regard are difficulty of navigational training tasks, the control interface and means of movement in the simulation (locomotion vs. teleportation), and the variety of environments encountered in training. VR training platforms with more natural movement and a greater variety of challenges might produce greater training outcomes beyond the target building. Future studies may also explore the does-repones relationships (number of sessions and length of each session), and directly compare outcomes between different digital training platforms and environmental designs, to better determine the best choices of activities and features.

Finally, in this study, we examined only the short-term effect of a single-session training, on a very direct target: wayfinding performance in a simulated environment. While the current study did not prespecify nor include cognitive functions as outcomes, effects on cognitive abilities related to wayfinding could be examined in a longitudinal study following regular exposure to training sessions over time.

In conclusion, this study indicates that VR wayfinding training is feasible with older adults and provides some evidence that this type of training  can effectively improve wayfinding performance in older adults. Although participants in the VR condition reported higher workload, there were no reported adverse effects. The effects of video training on performance were not found to be significantly different from control and were estimated to be slightly worse than VR condition with a very small effect size. Future studies should include larger samples and include participants who have challenges with wayfinding, explore different VR training design features and examine the long-term effect of simulated wayfinding training.

## Tables and Figures

**Table 1.** Baseline and Outcome Measures

| Construct | Measures | Reference | Theoretical Range | Time of Measurement |
|---|---|---|---|---|
| **Baseline Measures** | | | | |
| **Sense of Direction** | Santa Barbara Sense of Direction (SBSOD) | (Hegarty et al., 2002) | [1, 7] | online demographic survey |
| **Cognitive Assessment** | the Montreal Cognitive Assessment (MoCA) | (Nasreddine et al., 2005) | [0, 30] | in lab, before experiment |
| **Computer Proficiency** | Computer Proficiency Questionnaire (CPQ-12) | (Boot et al., 2015) | [6, 30][a] | online demographic survey |
| **Mobile Device Proficiency** | Mobile Device Proficiency Questionnaire (MDPQ-16) | (Roque & Boot, 2018) | [8, 40][a] | online demographic survey |
| **Feasibility Measures** | | | | |
| **Spatial Presence** | Self Location (SPSL) 4-item 5-point Likert Scale | (Vorderer et al., 2004) | [1, 5] | after intervention |
| | Possible Actions (SPPA) 4-item 5-point Likert Scale | (Vorderer et al., 2004) | [1, 5] | after intervention |
| **User Experience** | SUS: 10-item 5-point Likert Scale | (Brooke, 1996) | [0, 100][a] | after intervention |
| | Motion Sickness: 16-item 4-point Likert survey | (Kennedy et al., 1993) | [0.00, 235.62][a] | after intervention |
| **Main Outcome** | | | | |
| **Wayfinding Performance** | Task Duration | (Ruddle & Lessels, 2006) | [0, 600][b] | timed by experimenters |
| | Distance Traveled | (Ruddle & Lessels, 2006) | [0, ∞] | from recorded trajectory |
| **Secondary Outcome** | | | | |
| **Wayfinding Experience** | Spatial Anxiety: 5-item 4-point scale | (Zsido et al., 2020) | [1, 4] | After each task |
| | Workload: 6-item 7-point Likert Scale | (Hart & Staveland, 2019) | [1, 7] | after each task |
| **Spatial Learning** | Pointing to Task Origin (Pointing Error) | (Schinazi et al., 2013) | [0, 180][c] | after each task |
| | Distance Estimation from Task Origin (Distance Error) | (Schinazi et al., 2013) | [0.0, 4.6][d] | after each task |

*Note:* (a) rescaled to theoretical ranges per authors' instruction. (b) The time limit for wayfinding tasks was set to 600 seconds. (c) Pointing error was measured in degrees difference from the actual direction. (d) For the accuracy of Distance Estimation we used the absolute log value of the ratio (participants' estimates / correct distance); the cap of 4.6 is a 100-times difference.

**Table 2.** Demographic, Baseline and Feasibility Measures by Condition Group

| Measures | VR (N=17) | Video (N=16) | Control (N=16) | Overall (N=49) | Difference[b] |
|---|---|---|---|---|---|
| **Age** | 70.35 (8.03) | 71.06 (6.44) | 72.56 (8.69) | 71.31 (7.68) | F=0.35, p=0.711 |
| **Sex (Female)** | 14 (82%) | 10 (62%) | 14 (88%) | 38 (78%) | p=0.266 |
| **Ethnicity**[a] | | | | | p=0.422 |
| **White** | 17 (100%) | 15 (94%) | 15 (94%) | 47 (96%) | |
| **Asian** | 0 (0%) | 0 (0%) | 1 (6%) | 1 (2%) | |
| **Other** | 0 (0%) | 1 (6%) | 0 (0%) | 1 (2%) | |
| **MoCA** | 26.29 (1.99) | 26.56 (1.93) | 26.31 (2.60) | 26.39 (2.15) | F=0.08, p=0.927 |
| **Sense of Direction** | 4.33 (1.05) | 4.99 (1.01) | 4.68 (1.02) | 4.66 (1.04) | F=1.69, p=0.196 |
| **Computer Proficiency** | 27.53 (2.20) | 26.59 (4.08) | 27.88 (1.88) | 27.34 (2.86) | F=0.85, p=0.433 |
| **Mobile Device Proficiency** | 33.88 (7.12) | 31.38 (6.95) | 32.59 (6.80) | 32.64 (6.89) | F=0.54, p=0.589 |
| **Simulator Sickness** | 13.86 (16.71) | 31.09 (29.44) | 10.99 (17.99) | 18.55 (23.33) | F=3.92, p=0.027 |
| **Usability**[c] | 63.82 (14.55) | 59.06 (15.30) | 70.00 (15.00) | 64.29 (15.30) | F=2.15, p=0.128 |
| **SS: Self-Location** | 3.71 (0.94) | 2.72 (1.54) | 2.41 (1.18) | 2.96 (1.34) | F=4.97, p=0.011 |
| **SS: Possible Actions** | 2.99 (0.96) | 2.33 (1.23) | 2.33 (1.40) | 2.56 (1.22) | F=1.65, p=0.203 |

*Note:* SS: Spatial Presence. (a) No participants reported as Black or African American, American Indian, or Alaska Native, or Native Hawaiian or Pacific Islander. (b) Difference between group. F-test (2, 46) for continuous variables; Fisher’s exact test for categorical variables. (c) Measured in all groups for intervention similarity.

**Table 3.** Effects of VR training and Video training, compared with control

| | Similar Tasks | | | | | New Tasks | | | | |
|---|---|---|---|---|---|---|---|---|---|---|
| | **VR** (N=85) | **Video** (N=80) | **Control** (N=80) | **VR vs. Control[a] (95% CI)** | **Video vs. Control[a] (95% CI)** | **VR (N=75)** | **Video (N=72)** | **Control (N=75)** | **VR vs. Control[a] (95% CI)** | **Video vs. Control[a] (95% CI)** |
| **Duration** | 262.67 (181.00) | 241.90 (169.98) | 235.16 (174.72) | 0.68 [0.40, 1.14] | 0.77 [0.45, 1.31] | 190.71 (137.93) | 203.38 (161.83) | 234.93 (173.97) | 1.71 [0.99, 2.96] | 1.45 [0.84, 2.52] |
| **Distance** | 174.69 (121.13) | 189.81 (135.22) | 177.99 (146.80) | 0.84 [0.54, 1.30] | 0.68 [0.43, 1.08] | 159.68 (112.77) | 166.92 (132.43) | 206.05 (138.04) | **2.03 [1.15, 3.60]** | 1.72 [0.96, 3.08] |
| **excluded[b]** | 5 (5.9%) | 4 (5.0%) | 3 (3.8%) | | | 4 (5.3%) | 1 (1.4%) | 11 (14.7%) | | |
| **Tasks Finished** | 72 (85%) | 73 (91%) | 74 (92%) | | | 72 (96%) | 65 (90%) | 68 (91%) | | |
| **Spatial Anxiety** | 1.18 (0.26) | 1.23 (0.41) | 1.15 (0.24) | 0.01 [-0.22, 0.24] | 0.09 [-0.14, 0.33] | 1.07 (0.16) | 1.08 (0.19) | 1.16 (0.31) | -0.09 [-0.28, 0.10] | -0.04 [-0.23, 0.14] |
| **Workload** | 2.92 (1.16) | 2.70 (1.21) | 2.17 (1.16) | **0.71 [0.01, 1.40]** | 0.57 [-0.13, 1.28] | 2.21 (1.04) | 2.10 (1.14) | 1.95 (1.06) | 0.22 [-0.40, 0.83] | 0.20 [-0.41, 0.81] |
| **Pointing Error** | 83.93 (47.67) | 92.44 (46.43) | 78.60 (55.74) | 5.86 [-12.05, 23.77] | 13.36 [-4.79, 31.50] | 69.45 (44.15) | 71.65 (47.55) | 67.83 (47.96) | 3.88 [-19.30, 27.06] | 1.61 [-21.53, 24.75] |
| **Distance Error** | 0.87 (0.65) | 0.61 (0.49) | 0.95 (0.98) | -0.11 [-0.58, 0.35] | -0.30 [-0.78, 0.17] | 0.97 (0.91) | 0.92 (0.79) | 1.08 (1.04) | -0.12 [-0.41, 0.18] | -0.14 [-0.44, 0.16] |

*Note:* Data is analyzed at task level. The "hazard" in this study was finishing the task hence desirable. Significant effects are shown in bold. (a) Hazard Ratio for primary outcome (Duration and Distance), mean differences for secondary outcome. The "hazard" in this study was finishing the task hence desirable. (b) Distances of trajectories with incorrect start/end points were calculated, models fitted without imputation or missing data.

**Table 4. Summarized f-test results and effect sizes of secondary outcome**

| Measure | Similar Tasks | New Tasks |
|---|---|---|
| **Spatial Anxiety** | $F(2, 45.00)=0.54$, $p=0.588$, $\eta^2=0.02$, $\omega^2<0.01$ | $F(2, 38.49)=0.71$, $p=0.500$, $\eta^2=0.04$, $\omega^2<0.01$ |
| **Workload** | $F(2, 45.00)=3.44$, $p=0.041$, $\eta^2=0.13$, $\omega^2=0.09$ | $F(2, 41.11)=0.47$, $p=0.627$, $\eta^2=0.02$, $\omega^2<0.01$ |
| **Pointing Error** | $F(2, 45.00)=1.59$, $p=0.214$, $\eta^2=0.07$, $\omega^2=0.02$ | $F(2, 39.92)=0.08$, $p=0.920$, $\eta^2<0.01$, $\omega^2<0.01$ |
| **Distance Error** | $F(2, 45.00)=1.24$, $p=0.298$, $\eta^2=0.05$, $\omega^2=0.01$ | $F(2, 41.25)=0.77$, $p=0.469$, $\eta^2=0.04$, $\omega^2<0.01$ |

**Figure 1.** Screenshots from the VR and Video conditions. Images **A–C** show the 2D video, which included blue lines on the floor to guide participants' attention to the route. The camera in the videos moved inevitably along these predefined routes to the task destinations. Images **D–F** show 2D captures from the immersive 3D virtual environment. The light blue lines shown in these images are movement trajectories actively defined by participants as they sought to complete wayfinding tasks.

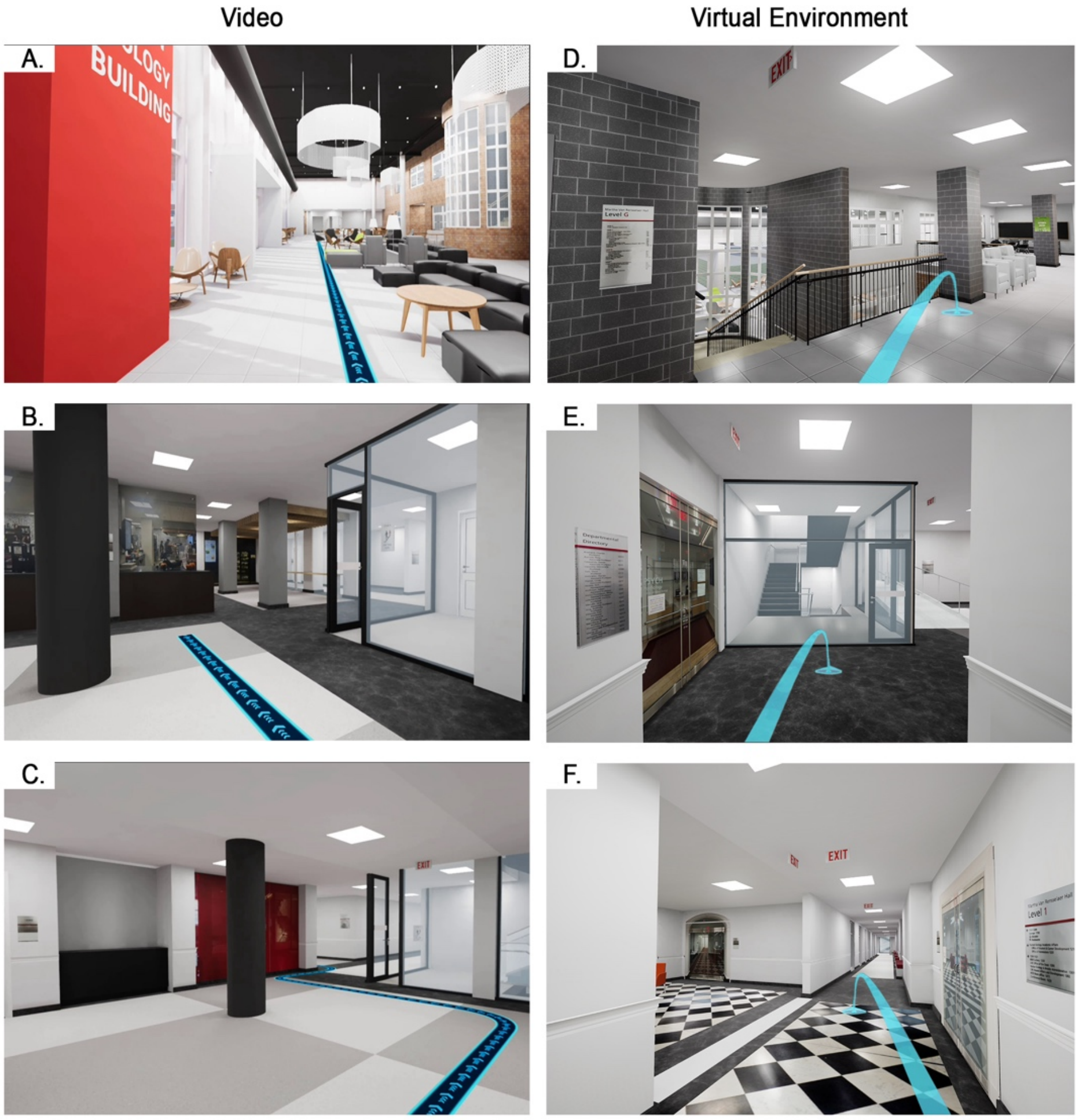

**Figure 2.** Wayfinding task success rates graphed against time spent on the tasks (A, B) and distance traveled during the tasks (C, D).

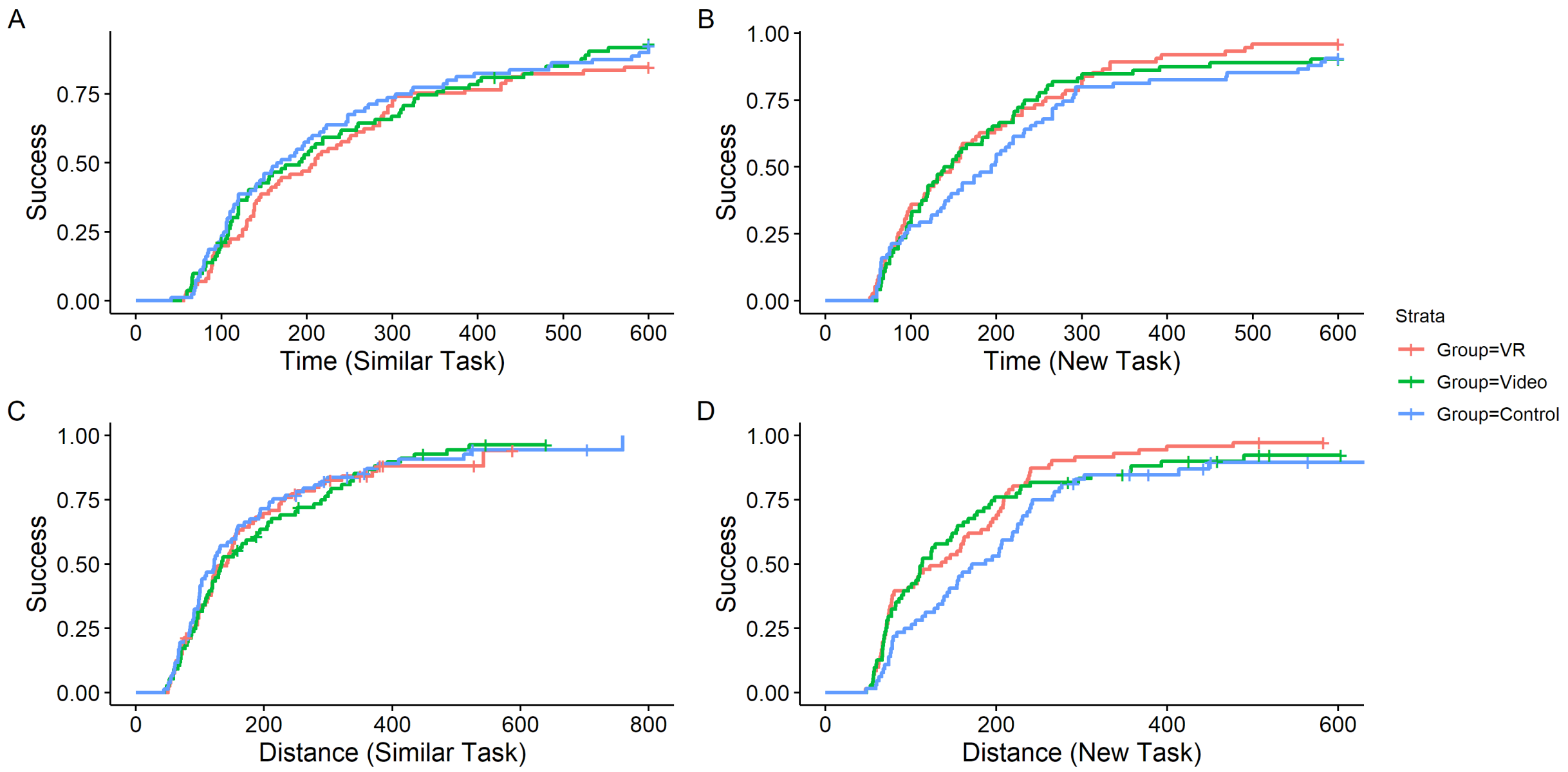

# Appendix

**Supplementary Figure S1.** Overview of the experiment procedure.

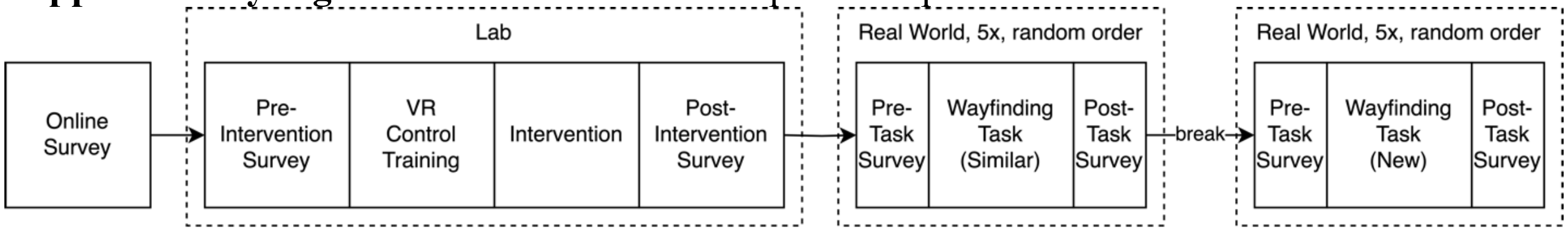

**Supplementary Table S1.** Summary of Wayfinding Tasks

| Task Number | Start Location | End Location | Change in Floor Level |
|---|---|---|---|
| 1 | T70 | Café | No (Single level) |
| 2 | Café | 1250 | Yes (Multi-Level) |
| 3 | 1250 | 1106 | No (Single level) |
| 4 | 1106 | G151 | Yes (Multi-Level) |
| 5 | G151 | T70 | Yes (Multi-Level) |
| 6 | G333 | 1300 | Yes (Multi-Level) |
| 7 | 1300 | 1210 | No (Single level) |
| 8 | 1210 | 1429 | No (Single level) |
| 9 | 1429 | T115 | Yes (Multi-Level) |
| 10 | T115 | G333 | Yes (Multi-Level) |

*Note:* Tasks 1–5 were closely similar to the intervention training; Tasks 6–10 took place in a part of the building that was not covered in the training.

**Supplementary Figure S2.** Floorplan view of the wayfinding tasks.

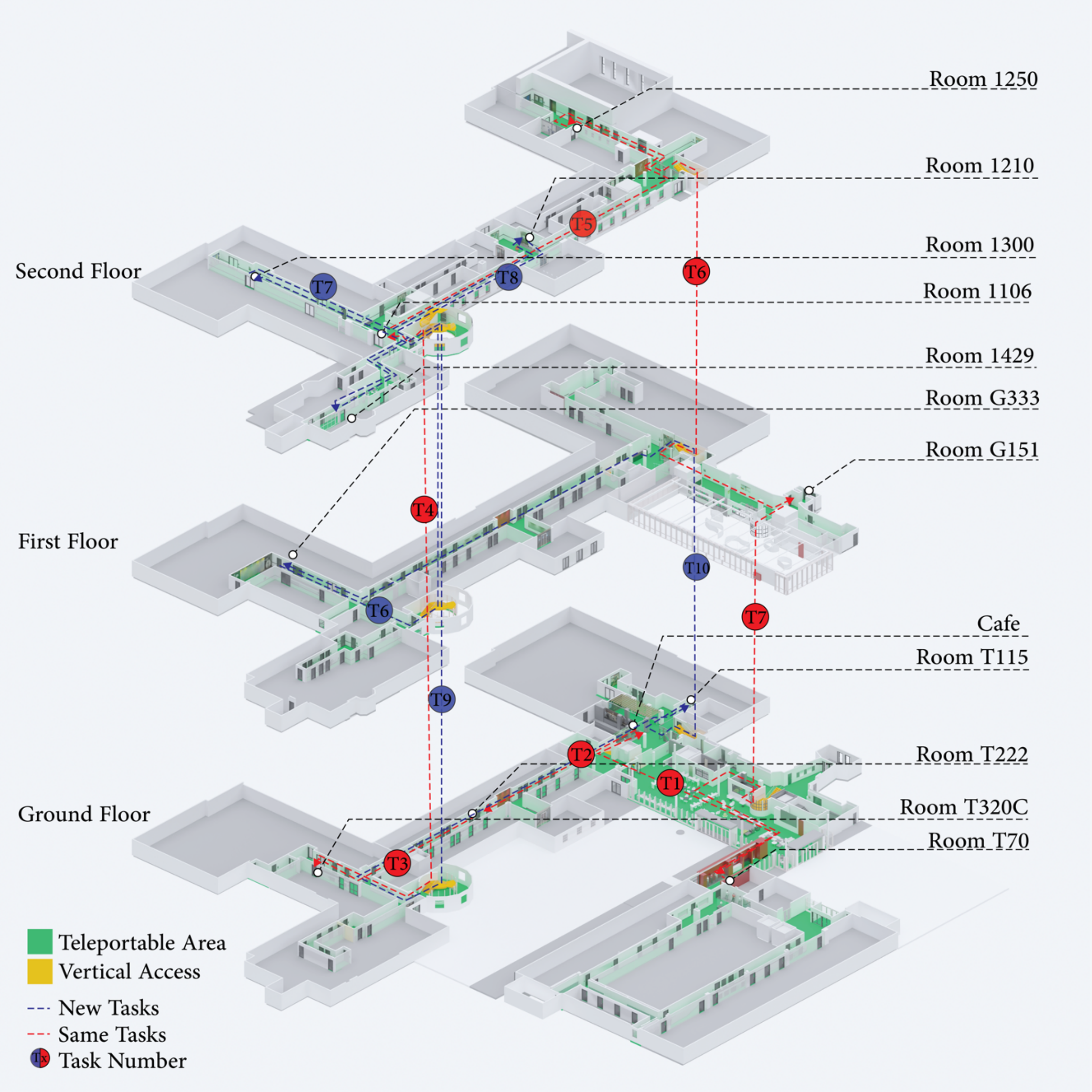